%% file: main.tex
\documentclass{article}
\usepackage{arxiv}
\usepackage[utf8]{inputenc} 
\usepackage[T1]{fontenc}    
\usepackage[hidelinks]{hyperref} 
\usepackage{url}            
\usepackage{booktabs}       
\usepackage{amsfonts}       
\usepackage{nicefrac}       
\usepackage{microtype}      
\usepackage{lipsum}
\usepackage{graphicx}
\usepackage{float}
\usepackage{amsmath}
\usepackage[version=3]{mhchem}
\usepackage{soul}
\usepackage{color}
\usepackage{authblk}
\usepackage{threeparttable}
\usepackage{tabularx}
\graphicspath{ {./Figures/} }
\usepackage{xcolor}
\usepackage{xr}
\usepackage{fontawesome5}

\newcommand{\etal}{\textit{et al.}}
\newcommand{\ie}{\textit{i.e.}}

\newcommand{\floridachem}{Department of Chemistry, University of Florida, Gainesville, FL 32611, USA}
\newcommand{\floridaqtp}{Quantum Theory Project, University of Florida, Gainesville, FL 32611, USA}
\newcommand{\floridamse}{Department of Materials Science \& Engineering, University of Florida, Gainesville, FL 32611, USA}

\newcommand{\nyuphysics}{Center for Soft Matter Research, Department of Physics, New York University, New York 10003, USA}
\newcommand{\nyusimons}{Simons Center for Computational Physical Chemistry, Department of Chemistry, New York University, New York 10003, USA}
\newcommand{\nyucourant}{Courant Institute of Mathematical Sciences, New York University, New York 10003, USA}
\newcommand{\nyucns}{Center for Neural Science, New York University, New York 10003, USA}
\newcommand{\minnesotacs}{Department of Computer Science \& Engineering, University of Minnesota, Minneapolis, MN 55455, USA}
\newcommand{\minnesotaaem}{Department of Aerospace Engineering and Mechanics, University of Minnesota, Minneapolis, MN 55455, USA}
\newcommand{\byu}{Department of Physics \& Astronomy, Brigham Young University, Provo, UT 84602, USA}

\begin{document}
\title{PropMolFlow: Property-Guided Molecule Generation with Geometry-Complete Flow Matching}

\author[1,2, 11]{Cheng Zeng}
\author[1,2, 11]{Jirui Jin }
\author[1,2]{Connor Ambrose}
\author[3]{George Karypis}
\author[4]{Mark Transtrum}
\author[5]{Ellad B. Tadmor}
\author[2, 6]{Richard G. Hennig}
\author[1,2]{Adrian Roitberg}
\author[7,8,9,10,*]{Stefano Martiniani}
\author[1,2,*]{Mingjie Liu}
\affil[1]{\floridachem}
\affil[2]{\floridaqtp}
\affil[3]{\minnesotacs}
\affil[4]{\byu}
\affil[5]{\minnesotaaem}
\affil[6]{\floridamse}
\affil[7]{\nyuphysics}
\affil[8]{\nyusimons}
\affil[9]{\nyucourant}
\affil[10]{\nyucns}
\affil[11]{These authors contribute equally: Cheng Zeng, Jirui Jin}
\affil[*]{e-mail: sm7683@nyu.edu, mingjieliu@ufl.edu}

\maketitle
\begin{abstract}
Molecule generation is advancing rapidly in chemical discovery and drug design. Flow matching methods have recently set the state of the art (SOTA) in unconditional molecule generation, surpassing score-based diffusion models. However, diffusion models still lead in property-guided generation. In this work, we introduce PropMolFlow, an approach for property-guided molecule generation based on geometry-complete SE(3)-equivariant flow matching. Integrating five different property embedding methods with a Gaussian expansion of scalar properties, PropMolFlow achieves competitive performance against previous SOTA diffusion models in conditional molecule generation while maintaining high structural stability and validity. Additionally, it enables faster sampling speed with fewer time steps compared to baseline models. We highlight the importance of validating the properties of generated molecules through DFT calculations. Furthermore, we introduce a task to assess the model’s ability to propose molecules with underrepresented property values, assessing its capacity for out-of-distribution generalization.
\end{abstract}

\section*{Introduction\label{sec:Introduction}}
\input{Sections/introduction}

\section*{Results and Discussion\label{sec:results}}
\input{Sections/results}

\subsection*{Discussion}
\input{Sections/discussion}

\section*{Methods\label{sec:methodology}}
\input{Sections/methodology}

\section*{Data availability}
All data can be found in the Zenodo repository~\cite{zenodo_propmolflow}, including revised QM9 SDF data, SDF files for generated raw molecules,  DFT-calculated structures and their molecular properties saved in the extxyz format, full property MAE and structural validity results in CSV format for all saved PropMolFlow checkpoint models, sampled structures and their full PoseBusters results for all baseline models, and EGNN property-predictor pretrained models and notebook examples to use the property predictors.

\section*{Code availability}
Our PropMolFlow implementation is available at https://github.com/Liu-Group-UF/PropMolFlow and on Zenodo~\cite{propmolflow_code}. 

\section*{Acknowledgments}
The authors acknowledge funding from NSF Grant OAC-2311632 and from the AI and Complex Computational Research Award of the University of Florida. S.M also acknowledges support from the Simons Center for Computational Physical Chemistry (Simons Foundation grant 839534). The authors also gracefully acknowledge UFIT Research Computing for computational resources and consultation, as well as the NVIDIA AI Technology Center at UF. We would like to thank Alex Morehead (University of Missouri) for providing the checkpoint models for the conditional generation of GeoLDM.

\bibliographystyle{unsrt} 
\bibliography{propmolflow}

\end{document}

%% file: Sections/introduction.tex
Deep generative models are promising to accelerate chemical discovery by statistically sampling molecular structures, reducing reliance on costly physics-based simulations~\cite{sanchez-lengeling2018}.
State-of-the-art (SOTA) 3D molecule generation primarily uses diffusion with equivariant graph neural networks (EGNN), enabling accurate sampling of molecular geometries.~\cite{hoogeboom2022, satorras2022, xu2023}.
Flow matching is an emerging alternative to diffusion models~\cite{liu2022, lipman2023, albergo2023}, offering flexibility over priors and probability paths while improving sampling efficiency across materials~\cite{hoellmer2025}, proteins~\cite{campbell2024}, and small molecules~\cite{dunn2024-2}.

Seminal diffusion~\cite{hoogeboom2022, xu2023} and flow-matching models~\cite{song2023} treat molecules as point clouds with continuous representations for discrete molecular modalities, and parameterize the generative process via E(3)-EGNN. However, continuous encodings disrespect inherently discrete features, such as atom types, and E(3) models cannot capture chirality~\cite{dumitrescu2024}. Recent studies also found bond orders in molecular graphs to be important in improving validity of generated molecules~\cite{vignac2023, dunn2024-2}. FlowMol~\cite{dunn2024-2} achieves SOTA unconditional molecule generation, by using geometry-complete SE(3) generative processes, explicit bond-order modeling, and discrete flow matching for atom types, formal charges and bond orders, though its extension to property-guided generation remains unaddressed.   

Property-guided molecule generation aims to design molecules that satisfy target properties, often by concatenating property values with node features in molecular graphs~\cite{hoogeboom2022, xu2023,  morehead2024} --- a strategy mainly explored on the single-molecule QM9 data~\cite{ramakrishnan2014quantum, wu2018moleculenet}. Although effective, this approach may oversimplify how properties interact with molecular structures; for instance, Gebauer et al.~\cite{gebauer2022} suggest Gaussian expansions can extract property information more robustly. Systematic studies of property embedding methods are lacking, and optimal strategies may vary by properties. To validate properties of generated molecules, prior methods use separate property predictors trained on the fully relaxed QM9 molecules, which may struggle to make accurate predictions on non-relaxed, generated molecules. Moreover, property-guided generation
is typically evaluated in an ‘in-distribution’ setting, where property values and atom counts sampled within the QM9
distribution are used to generate molecules~\cite{hoogeboom2022}. Chemical discovery requires generating molecules with underrepresented or out-of-distribution (OOD) property values.  

We introduce PropMolFlow, a property-guided molecule generation framework that integrates various property embedding methods with an SE(3)-equivariant flow-matching process~\cite{dunn2024-2}. 
We evaluate PropMolFlow on QM9, demonstrating its competitive performance versus previous methods, while achieving faster inference.
Besides, we propose an out-of-distribution generation task, and validate properties of generated molecules using extensive density functional theory (DFT) calculations. Finally, we introduce the close-shell ratio and revised stability metrics to complement existing metrics and ensure more comprehensive evaluations for molecule generation.

%% file: Sections/results.tex
\newcommand{\cs}{Concatenate\_Sum~}
\newcommand{\cm}{Concatenate\_Multiply~}
\newcommand{\concat}{Concatenate~}
\newcommand{\su}{Sum~}
\newcommand{\m}{Multiply~}

\subsection*{Overview of PropMolFlow}

\begin{figure}
    \centering
    \includegraphics[width=0.95\linewidth]{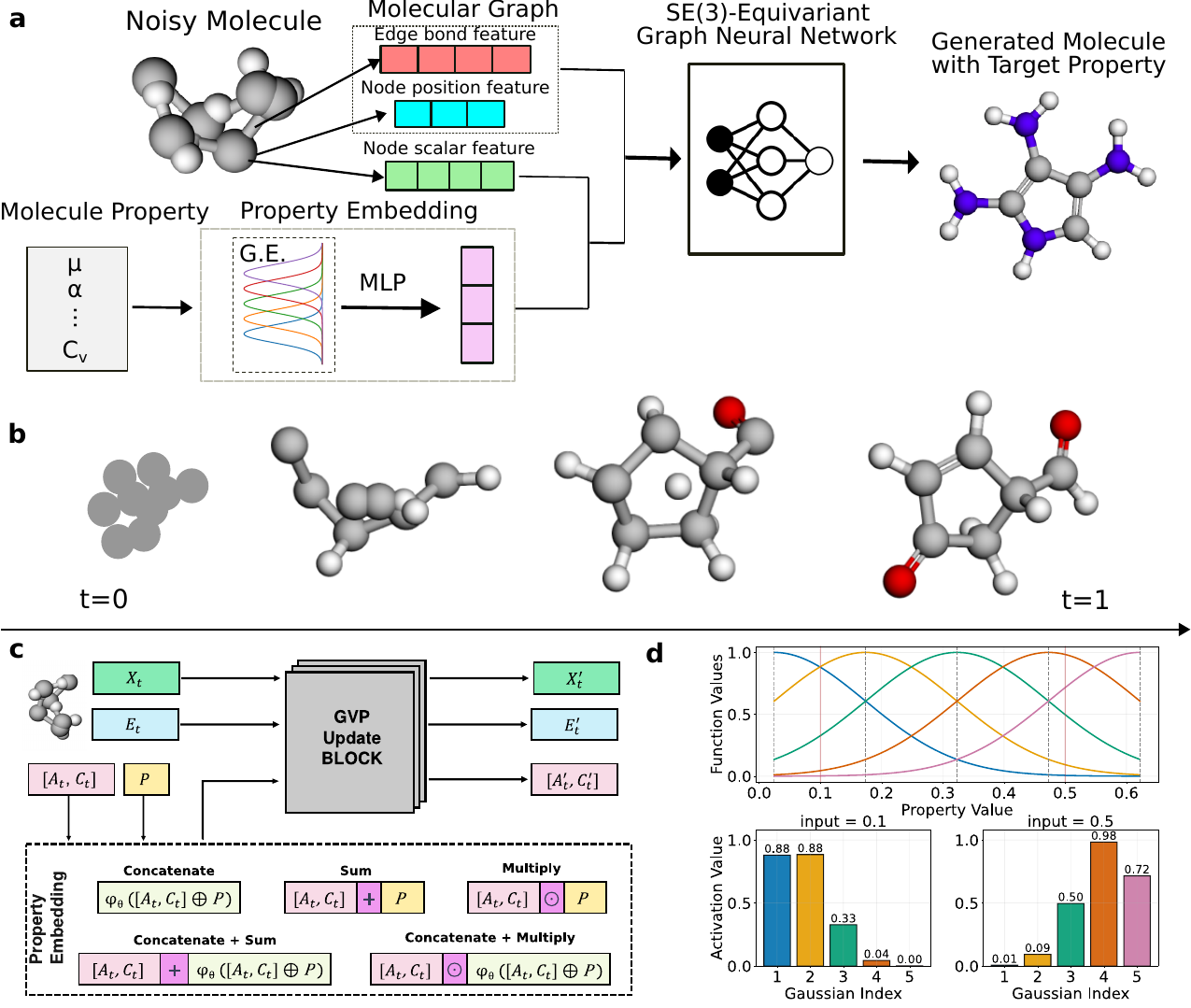}
    \caption{Overview of the PropMolFlow methodology. \textbf{a}, PropMolFlow models are jointly trained on a molecular graph and property embedding. A molecular graph includes node scalar features, node position features and edge bond features. A property embedding comprises an optional Gaussian expansion mapping (G.E.) followed by a multilayer perceptron (MLP) that projects a scalar property to a high-dimensional embedding space. Conditioning on the property is achieved by the interaction between the property embedding and node scalar features. \textbf{b}, A joint flow matching process is used to generate different molecule modalities together. \textbf{c}, Five interaction types between a property embedding `P' and a molecular graph. Node scalar features $[A_t, C_t]$ include atom type $A$ and formal charge $C$ at time $t$. An MLP transformation $\varphi_\theta$ is applied to convert the dimension back to that of the original $[A_t, C_t]$, where necessary. `$\odot$' represents an element-wise Hadamard product, and `$\oplus$' indicates a `Concatenate' operation. $E_t$ and $X_t$ represent respective bond edge features and node position features.  \textbf{d}, Gaussian expansion as augmented property embedding. Curves in the top panel correspond to five Gaussian basis functions that are evenly spaced between the minimum and maximum property values. Centers of Gaussians are marked by gray dashed lines, and the red solid lines represent two example inputs. The bottom panels show the function values for each Gaussian for two inputs. In a molecule configuration, white, gray, red, and blue indicate H, C, O, and N, respectively.}
    \label{fig:propmolflow_overview}
\end{figure}

PropMolFlow builds on top of the FlowMol architecture.~\cite{dunn2024-2} 
The model generates samples by integrating over NN-parameterized conditional velocity fields~\cite{lipman2023, albergo2023, liu2022}.
An SE(3)-EGNN based on geometric vector perceptrons (GVP)~\cite{jing2021} is used and molecules are represented by fully-connected graphs (Figure \ref{fig:propmolflow_overview}a).
The generation is achieved by interacting property embeddings and node scalar features.
Property embeddings are constructed by a non-trainable Gaussian expansion, followed with a shallow multilayer perceptron (MLP).
This Gaussian expansion is optional, and its utility depends on property types and evaluation tasks.
Once trained, the NN defines a joint flow matching that denoises the molecular graph by simultaneously updating all modalities---such as atom types, bond orders, and 3D coordinates---to generate molecules (Figure \ref{fig:propmolflow_overview}b).
Details of flow matching and neural networks are provided in the \hyperref[sec:joint_fm]{`Joint Flow Matching'} and \hyperref[sec:model_architecture]{`Model Architecture'} sections. 

Five embedding methods and usage of Gaussian expansion are explored (Figure \ref{fig:propmolflow_overview}c,d). 
We evaluate the property-guided generation over six molecular properties in QM9.
Details for property embeddings are provided in the \hyperref[sec:prop_emb]{`Property Embedding Operations'} and \hyperref[sec:ge]{`Gaussian Expansion'} sections.
PropMolFlow is trained on a revised QM9 SDF data with explicit hydrogen atoms, bond orders and formal charges~\cite{ramakrishnan2014quantum, wu2018moleculenet}.
This revised SDF file fixed the bond and charge inconsistencies of the original SDF file distributed by DeepChem~\cite{wu2018moleculenet}.
Data details are provided in the \hyperref[sec:dataset]{`Data'} section.

\subsection*{Competitive property-guided generation with PropMolFlow}\label{sec:propmolflow_results}
We evaluated PropMolFlow on its ability to generate molecules with target properties, molecular structural validity, and  inference speeds. These assessments were conducted under the in-distribution (ID) task.
This task assesses the model's capability to generate structures spanning a spectrum of joint target properties and atom counts within data distributions and has been adopted in prior studies~\cite{xu2023,song2023,morehead2024,bao2023}.  
Generative models sample molecules by property inputs, and a separate property predictor evaluates properties of generated molecules.
Predicted values are compared to input values, and the mean absolute error (MAE) was used as the evaluation metric.
A lower MAE indicates better model performance.
Further details of how to sample molecules and GVP regressors are provided in the \hyperref[sec:inference]{`Inference'} and \hyperref[sec:gvp_details]{`GVP regressor details'} sections.

We compared PropMolFlow with five baseline models, including EEGSDE~\cite{bao2023}, EquiFM~\cite{song2023}, GeoLDM~\cite{xu2023}, GCDM~\cite{morehead2024} and JODO~\cite{huang2024}, all trained on QM9. 
Two intuitive baselines are also included to represent the respective upper and lower bounds of MAEs, namely ``Random (Upper-bound)'' and ``QM9 (Lower-bound)'', along with another baseline ``\# Atoms'' (see the \hyperref[sec:baselines]{`Baselines'} section for more details).

\begin{table}
    \centering
    \caption{\textbf{Performance of PropMolFlow with respect to property alignment.}
    Results for the baseline models use an equivariant graph neural network (EGNN) as the regressor,
    whereas PropMolFlow results use our pretrained SE(3) GVP regressor as well as an EGNN regressor.
    JODO results were reported on our sampled molecules using publicly available model checkpoints.}
    \begin{tabular}{c|cccccc}
    \hline
    Property & $\alpha$ & $\Delta \epsilon$ & $\epsilon_{\rm{HOMO}}$ & $\epsilon_{\rm{LUMO}}$ & $\mu$ & $C_v$ \\
    Units & Bohr$^3$ & meV & meV & meV & Debye & cal/(mol$\cdot$K) \\
    \hline
    QM9 (Lower-bound)       & 0.10 & 64   & 39  & 36  & 0.043 & 0.040 \\
    Random (Upper-bound)    & 9.01 & 1470 & 645 & 1457 & 1.616 & 6.857 \\
    \# Atoms                & 3.86 & 866  & 426 & 813  & 1.053 & 1.971 \\
    \hline
    EEGSDE                  & 2.62 & 542 & 302 & 496 & 0.858 & 1.037 \\
    EquiFM                  & 2.41 & 591 & 337 & 530 & 1.106 & 1.033 \\
    GeoLDM                  & 2.37 & 587 & 340 & 522 & 1.108 & 1.025 \\
    GCDM                    & 1.97 & 602 & 344 & 479 & 0.844 & 0.689 \\
    JODO                    & 1.44 & 333 & 231 & 260 & 0.620 & 0.580 \\
    PropMolFlow, GVP        & 1.31 & 391 & 254 & 315 & 0.620 & 0.626 \\
    PropMolFlow, EGNN       & 1.36 & 391 & 246 & 312 & 0.632 & 0.640 \\
    \hline
    \end{tabular}
    \label{tab:id}
\end{table}

Overall, PropMolFlow achieves competitive performance against the SOTA models (Table \ref{tab:id}).
It shows the lowest MAE for $\alpha$ and $\mu$, and comparable performance for $C_v$ and $\epsilon_{\rm{HOMO}}$ versus the SOTA JODO, and the second best performance for $\epsilon_{\rm{LUMO}}$ and $\Delta \epsilon$.
The superior performance of PropMolFlow benefits from the thorough exploration of property-dependent optimal embedding methods (Supplementary Table \ref{tab:mae_range} and Supplementary Section \ref{sec:model_stats}).
For instance, the MAE for $\Delta \epsilon$ can vary by more than 40\%, from 391 meV using a `Concatenate\_Sum' embedding to 551 meV using a `Concatenate' embedding, both without Gaussian expansions.
To avoid bias of comparison using different predictors, we also report MAEs using EGNN predictors released by Bao et al. ~\cite{bao2023} in Table~\ref{tab:id}.
It confirms that different predictors produce consistent MAEs.

\subsection*{Rapid molecule generation with high structural fidelity}\label{sec:structural_validity}

Structural validity and inference speeds are critical in success of chemical discovery with generative models.
Structural validity refers to whether generated molecules conform to chemical rules. This was measured using five metrics, including atomic stability, molecule stability, RDKit validity~\cite{rdkit}, `uniqueness and validity', PoseBusters validity~\cite{buttenschoen2024}, and closed-shell ratio.
Inference speeds are evaluated by the wall-clock time for generating 10,000 molecules.
Details of structural metrics and computational settings for sampling are provided in respective \hyperref[sec:evaluation_metrics]{`Evaluation metrics'} and \hyperref[sec:computational_setting]{`Computational settings'} section.
\begin{figure}[htbp]
  \centering
  \includegraphics[width=0.98\linewidth]{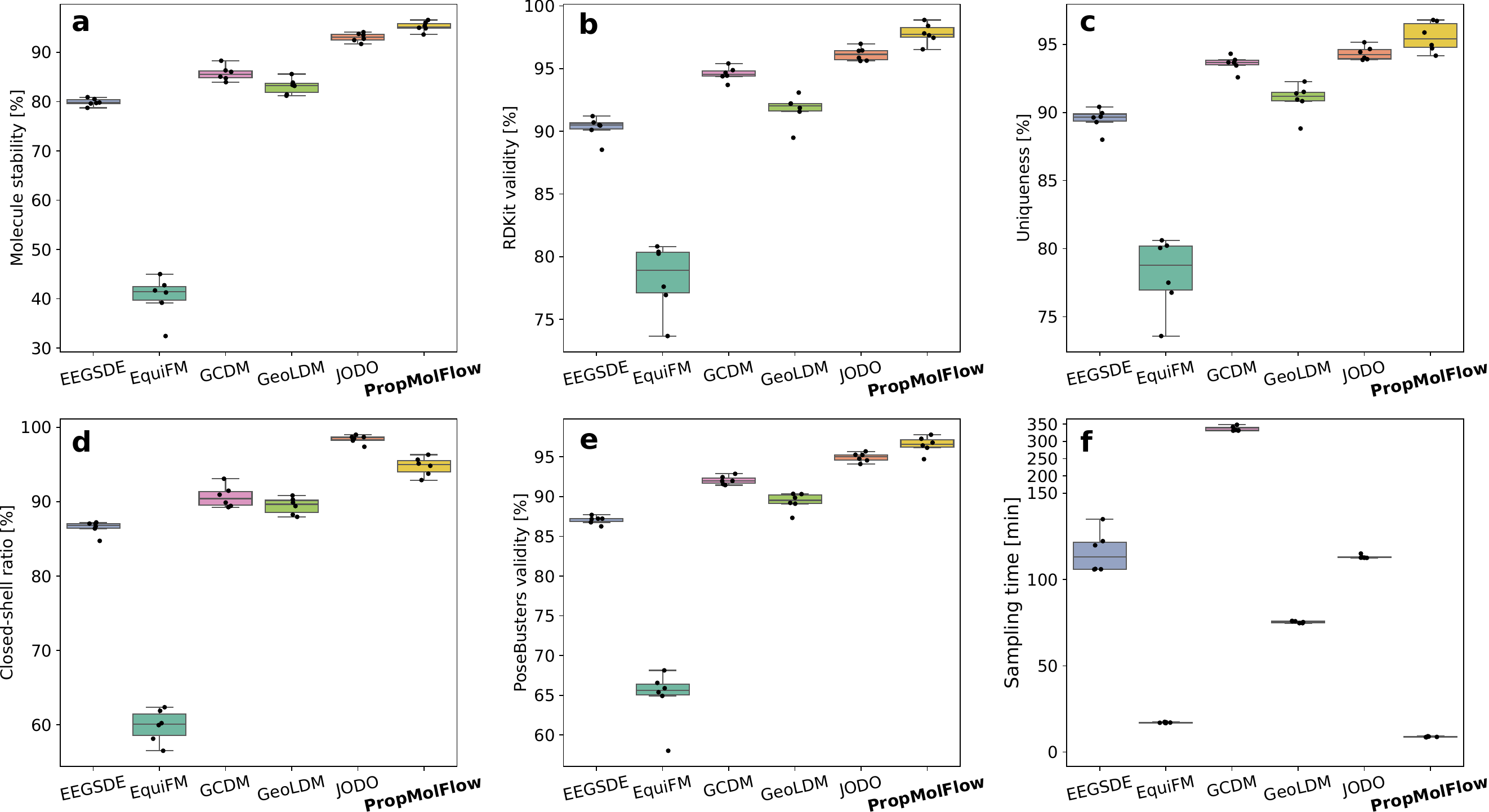}
  \caption{\textbf{Chemical validity and sampling efficiency of PropMolFlow against five baseline models.}
  \textbf{a,} Molecule stability.
  \textbf{b,} RDKit validity.
  \textbf{c,} Uniqueness.
  \textbf{d,} Closed-shell ratio.
  \textbf{e,} PoseBusters validity.
  \textbf{f,} Sampling time.
  The y-axis for sampling time uses a broken scale to expand 0--150 min and compress 150--360 min by a ratio of 5 for visual clarity.
  Each box plot summarizes the metric values computed for six molecular properties ($n=6$) and 10,000 sampled molecules.
  The median is shown as a solid line.
  The edges of the box correspond to the first and third quartiles, and the whiskers extend to values within 1.5$\times$ interquartile ranges.
  All individual data points are overlaid as black dots.
  PropMolFlow results use the top-performing models of each property in the ID tasks for property alignment.}
  \label{fig:standard_metrics}
\end{figure}

PropMolFlow consistently outperforms baseline models  across all structural metrics, except for closed-shell ratios, where it slightly underperforms JODO (Fig.~\ref{fig:standard_metrics}a--e and Supplementary Table.~\ref{tab:standard_metrics}).
Flow matching is normally much faster than diffusion models due to its shorter, deterministic probability paths and optimal transport (Supplementary Section \ref{sec:ot})~\cite{song2023}. 
PropMolFlow requires only 100 time steps against 1000 steps for diffusion models, hence a speedup at least 8$\times$ over diffusion-based models and nearly 2$\times$ faster than EquiFM (Fig.~\ref{fig:standard_metrics}f).
We also report per-property comparisons in Supplementary Section~\ref{sec:si_results}.
To further assess novelty of generated molecules, Supplementary Figure~\ref{fig:si_id_tanimoto} shows the maximum Tanimoto similarity of generated molecules to the training set using Morgan fingerprints~\cite{morgan1965}.
Applying a cutoff of 0.8, novelty ratios are 66--72\% across all properties, indicating the model's ability to produce a substantial number of previously unseen molecules.

\subsection*{Systematic inductive bias of property predictors}\label{sec:gvp_reliability}

\begin{figure}[htbp]
  \centering
  \includegraphics[width=0.98\linewidth]{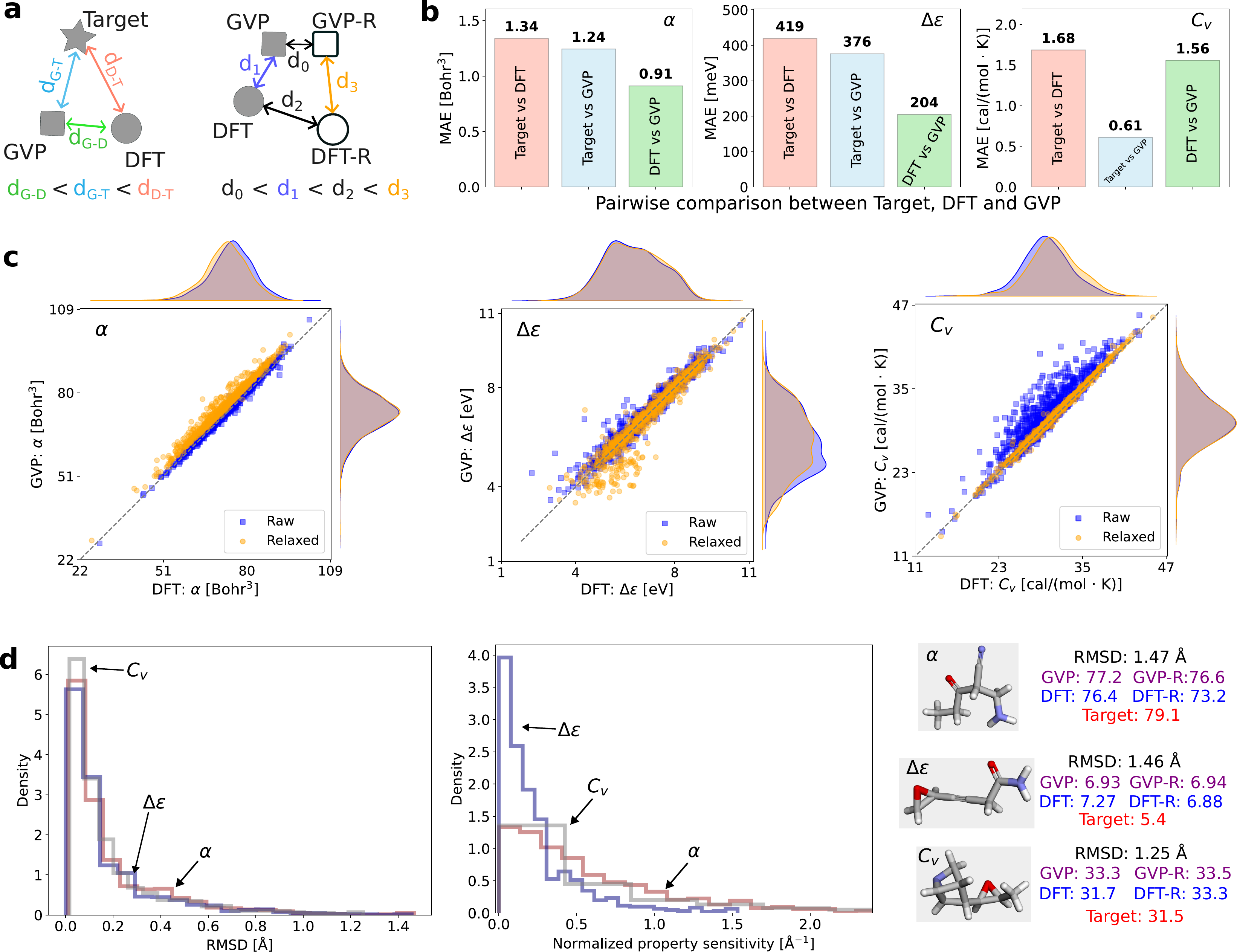}
  \caption{\textbf{Performance of GVP property predictors without and with DFT relaxation.}
  \textbf{a,} Comparison between Target, DFT and GVP shows the reliability of GVP in evaluating MAE metrics commonly used for property-guided generation.
  Comparison between GVP and GVP-R, and between DFT and DFT-R shows the structural dependence of GVP-predicted and DFT-predicted properties, respectively.
  Comparison between GVP and DFT with and without relaxation shows the reliability of GVP in capturing ground-truth DFT values for both raw and relaxed structures.
  `Target', `DFT' and `GVP' denote input, DFT-calculated and GVP-predicted property values on raw molecules, respectively.
  `DFT-R' and `GVP-R' refer to values evaluated on DFT-relaxed molecules.
  The $d$ indicates the MAE distances between two property-value vectors.
  \textbf{b,} Pairwise comparison between Target, DFT and GVP on raw molecules.
  \textbf{c,} GVP versus DFT for both raw and DFT-relaxed molecules.
  \textbf{d,} Root mean squared distances (RMSDs) and normalized property sensitivity due to DFT relaxation.
  Molecules with the highest RMSDs for each property and their corresponding DFT and GVP values are shown.
  In molecular representations, gray, red, blue and white indicate C, O, N and H atoms, respectively.
  Property values for $\alpha$, $\Delta\epsilon$ and $C_v$ are in units of Bohr$^3$, eV and cal\,mol$^{-1}$\,K$^{-1}$, respectively.}
  \label{fig:target_gvp_dft}
\end{figure}

A property predictor is needed to evaluate the generated molecules. Typically,
this predictor shares the same architecture as the generative model, which may introduce inductive biases in predictions.
In this work, DFT calculations were carried out on selected and filtered molecules and compared to input (`Target') and GVP-predicted values, without and with (`-R') structural relaxations.
For example, we show orders of pairwise MAE distances for $\alpha$ in Fig.~\ref{fig:target_gvp_dft}a.
We show results for $\alpha$, $\Delta \epsilon$ and $C_v$ in Fig.~\ref{fig:target_gvp_dft}b--d, and for the other properties in Supplementary Figure \ref{fig:si_target_dft_gvp}, which are similar to $\Delta \epsilon$.
DFT settings are included in the \hyperref[sec:dft_details]{`DFT'} section,
the filtering procedure are provided in the \hyperref[sec:molecule_filtering]{`Molecule filtering'} section, and numbers of filtered molecules are listed in Supplementary Table \ref{tab:no_sp_dft}.

First, we examine GVP on the raw generated molecules (Fig.~\ref{fig:target_gvp_dft}b).
MAEs of ``Target vs. DFT'' are comparable to ``Target vs. GVP'' for $\alpha$ and $\Delta \epsilon$, suggesting that GVP predictors are reliable for estimating ``Target vs. DFT'' MAEs.
However, GVP consistently underestimates the MAEs, revealing an inductive bias.
For $C_v$, the ``Target vs. DFT'' MAE is much larger than ``Target vs. GVP''.
This is because DFT-computed $C_v$ depends on vibrational frequencies, which are highly sensitive to geometry. 
Relaxation resolves this issue, reducing the MAE to 0.68 cal/(mol $\cdot$ K), still higher than, but closer to ``Target vs. GVP'' (0.61).
For chemical discovery, it is also important to relax the structures.
Fig.~\ref{fig:target_gvp_dft}c shows a much closer agreement between DFT and GVP after relaxation for $C_v$, while a notable offset is observed for $\alpha$.
It indicates that GVP is more reliable on unrelaxed structures, perhaps due to its inductive bias. 
This is also true for other properties (Supplementary Figure \ref{fig:id_mae_heatmap}).

To further quantify the relaxation effect, we calculated the root mean square distances (RMSDs) between atomic positions before and after relaxation. 
Fig.~\ref{fig:target_gvp_dft}d shows that RMSD distributions are left-skewed toward 0 \AA, meaning most unrelaxed structures are close to their relaxed counterparts.
To estimate the property sensitivity to relaxation, we calculated the normalized property sensitivity to allow comparisons between different properties.
Definitions of RMSDs and normalized property sensitivity are provided in the \hyperref[sec:evaluation_metrics]{`Evaluation metrics'} section.
Despite similar RMSD distributions, the differences are more pronounced in the property sensitivity, following the order of $C_v > \alpha > \Delta \epsilon$.
Fig.~\ref{fig:target_gvp_dft}d also shows configurations with the highest RMSDs, their target, DFT-calculated, and GVP-predicted values.
DFT values with relaxation can be either closer or farther to Target.
Despite high RMSDs, GVP values are nearly unaltered with relaxation, showing a much weaker structural dependence of GVP \textit{versus} DFT.
Overall, GVP predictions are close to DFT values for all cases, hence it can offer a statistically reliable evaluation of PropMolFlow models despite the inductive biases.

\subsection*{Interpolated structures aligned with chemical intuition}\label{sec:interpolation}
To assess whether PropMolFlow has learned a smooth structure--property relationship, we performed interpolation by varying target values while fixing atom counts of molecules at 19.
For each value, ten molecules were sampled, and the one whose DFT-calculated property is closest to the input was selected. 
We show the configurations for selected molecules, and minimum and maximum target values of each property along with corresponding DFT values in Fig.~\ref{fig:interpolation}. Numeric results for all configurations are provided in Supplementary Table \ref{tab:interpolation}.

\begin{figure}[htbp]
  \centering
  \includegraphics[width=0.95\linewidth]{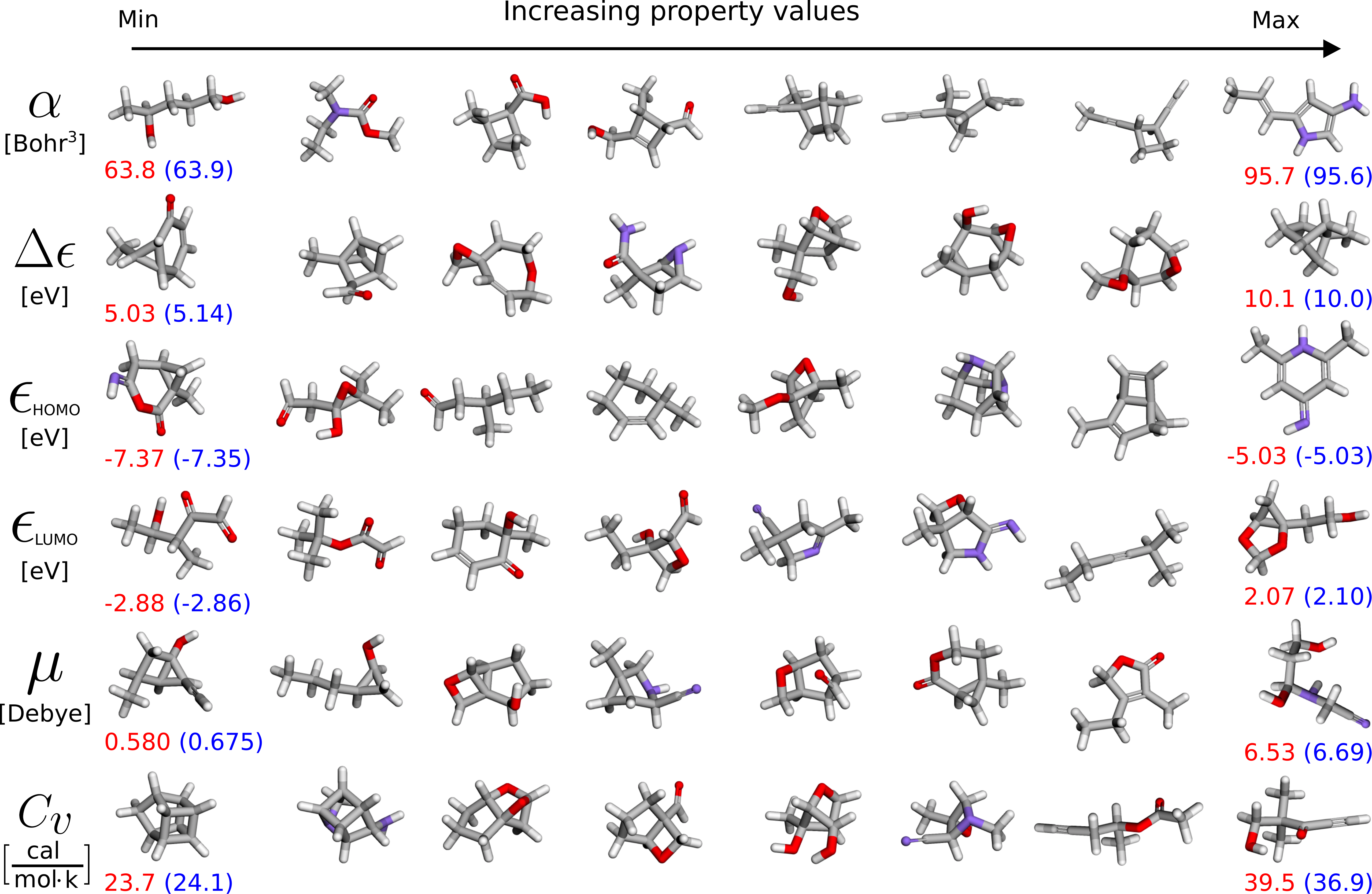}
  \caption{\textbf{Interpolation study by varying property values.}
  The minimum and maximum target properties (red) and the corresponding DFT-calculated properties (blue) are shown below each configuration.
  All molecules shown pass the filtering criteria and have DFT values closest to the target properties.
  In molecular representations, gray, white, red and blue indicate C, H, O and N atoms, respectively.
  Property units are provided in square brackets under each property symbol.}
  \label{fig:interpolation}
\end{figure}

The interpolated molecules exhibit chemically intuitive structural changes when target values increase. 
For instance, higher $\alpha$ values largely correspond to more elongated molecular shapes.
Higher $\mu$ and $C_v$ tend to show respective polar functional groups (-C$\equiv$N and -COO) and extended structures with reduced steric hindrance. 
High HOMO-LUMO gap, HOMO, LUMO energies correspond to more polarized $\sigma$ bonds, $\pi$ conjugated systems, and saturated $\sigma$ bonds, respectively.
Yet, the generated molecule may not correspond to the most stable isomer, probably due to data limitations or sampling stochasticity. 
For example, conditioned on HOMO, the generated `2,6-Dimethyl-1H-pyridin-4-imine' is likely less stable than its tautomer `5-Amino-3-methyl-2H-pyridin-4-imine'. 
2D molecular graph, SMILES and energy comparison of these two isomers are provided in Supplementary Figure~\ref{fig:tautomer_compare}. 

\subsection*{Toward out-of-distribution generation}\label{sec:ood_gen}

\begin{figure}[htbp]
  \centering
  \includegraphics[width=0.95\linewidth]{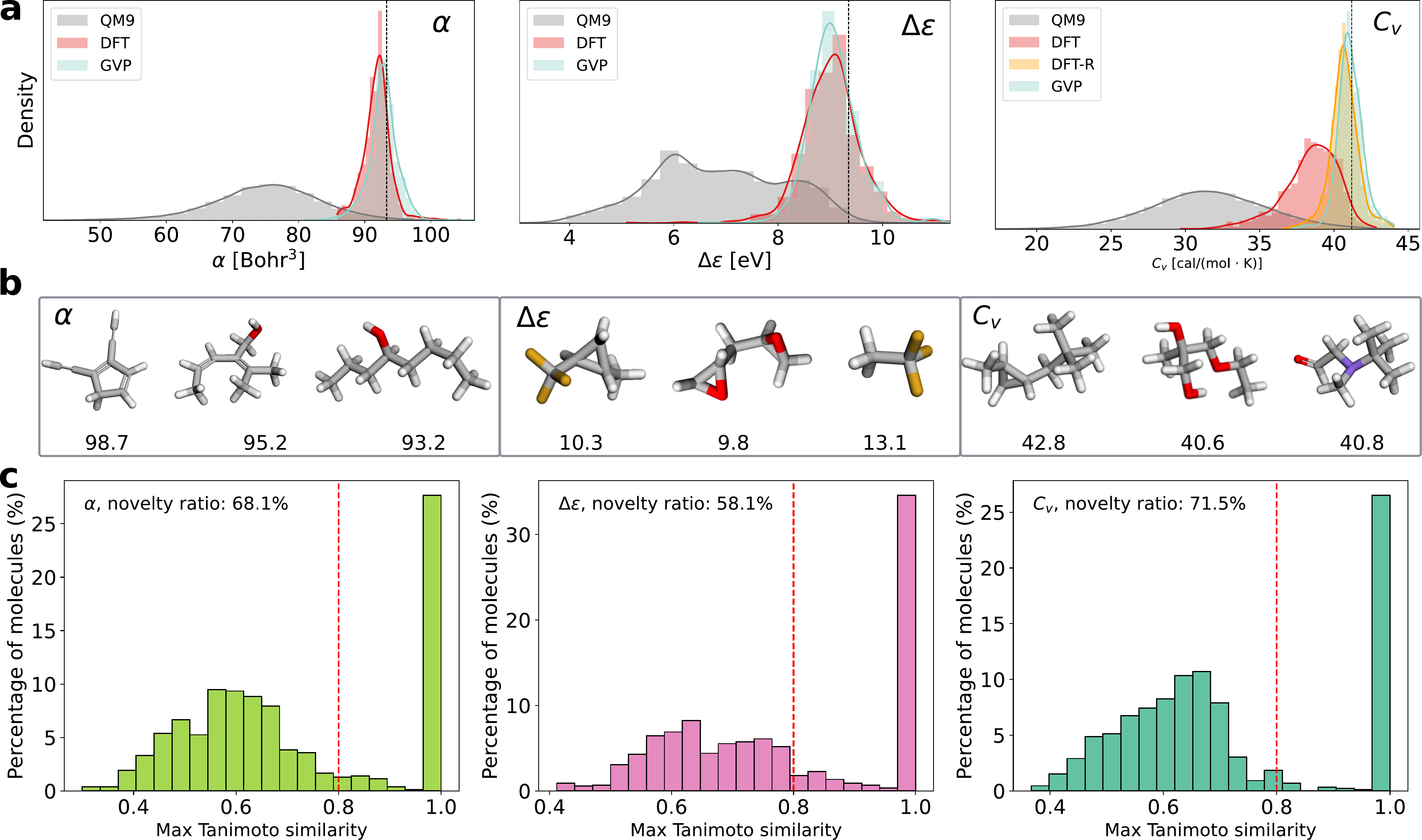}
  \caption{\textbf{Toward out-of-distribution generation.}
  \textbf{a,} Distribution of DFT-calculated and GVP-predicted property values for molecules generated by PropMolFlow; the property distribution of the QM9 training data is also shown.
  The vertical black dashed line in the histograms denotes the target property value $q_{0.99}$, corresponding to the 99th percentile of the training-data distribution.
  Curves overlaid on the histograms are kernel density estimation fits.
  \textbf{b,} Three example molecules absent from QM9 but present in a larger PubChem dataset are shown in the left panel.
  Numbers below the configurations indicate DFT-calculated property values for raw molecules generated by PropMolFlow.
  In molecular representations, gray, white, red, blue and yellow indicate C, H, O, N and F atoms, respectively.
  Property values for $\alpha$, $\Delta\epsilon$ and $C_v$ are in units of Bohr$^3$, eV and cal\,mol$^{-1}$\,K$^{-1}$, respectively.
  \textbf{c,} Maximum Tanimoto similarity between generated, filtered molecules and the training data computed using Morgan fingerprints.
  Dashed lines indicate the 0.8 similarity cutoff used to define novel molecules.}
  \label{fig:ood}
\end{figure}

We propose an OOD task to test generation for underrepresented property values.
The 99th quantile of the training data distribution was chosen as the target value and the atom counts were chosen according to the \hyperref[sec:inference]{`Inference'} section.
Specific target values are provided in Supplementary Table \ref{tab:si_prop_val_stats}.
1,000 molecules were sampled with top-performing models in this task (Supplementary Table \ref{tab:best_model}), and we filtered structures using the same procedure as that used in the ID task.
We compare DFT and GVP property distributions against that of the QM9 training data (Fig.~\ref{fig:ood}a).
DFT distributions shift toward the targets (dashed vertical lines). 
The distribution mode is very close to the target for $\alpha$. 
For $\Delta \epsilon$ and $C_v$, DFT distribution modes are off the targets by a small margin, for example 0.41 eV for $\Delta \epsilon$ (8.95 versus target 9.36 eV).
This deviation is likely due to the scarcity of training data in the corresponding property region; when conditioning on the 50th quantile (medians), modes are closer to the targets (Supplementary Figure~\ref{fig:si_q05_vs_q099}).
Moreover, GVP distributions are almost overlapped with DFT ones, suggesting their desired extrapolation performance.

To examine the novelty of generated molecules, Fig.~\ref{fig:ood}b presents three generated molecules whose SMILES representations are not part of QM9 but are found in the larger PubChem database~\cite{kim2021}.
We also evaluated the novelty of generated molecules by Tanimoto similarity to the training data. Fig.~\ref{fig:ood}c indicates that 20--35\% of the molecules exist in the training data.
Using 0.8 as a cutoff, novelty ratios are 58--75\% across all properties, demonstrating substantial exploration beyond the training data distribution.
Results for other properties are provided in Supplementary Figure \ref{fig:si_ood}, and additional extrapolation to more extreme property values or atom counts can be found in respective Supplementary Figure~\ref{fig:si_various_values} and Supplementary Figure~\ref{fig:si_ood_no_atoms}.

%% file: Sections/discussion.tex
While PropMolFlow can generate chemically valid and novel molecules with strong property alignment and fast inference, its ability to extrapolate could be further enhanced, potentially by integrating the model into an active-learning or reinforcement-learning framework. The current implementation is limited to the small-molecule QM9 dataset; extending PropMolFlow to larger molecular datasets and conditioning on properties of greater practical relevance represents an important future direction.
In some cases, the property distribution of generated molecules exhibits deviations from target values. Property alignment could be improved by incorporating advanced guidance mechanisms with tunable guidance weights~\cite{ho2022, karras2024}. Moreover, despite faster inference than diffusion models, flow matching still requires multi-step integration . Additional efficiency gains may be possible by learning flow maps that support direct jumps along probability paths~\cite{boffi2025}. However, these approaches have shown effectiveness only on continuous data; learning flow maps for multi-modal 3D molecules remains a challenge.

Although extending PropMolFlow to multiple properties is conceptually straightforward, the optimal strategy for synthesizing different properties and for managing potential correlations remains unclear. Another challenge concerns conformational stability: PropMolFlow does not guarantee the generation of the most stable conformers. Injecting energies into \textit{de novo} generation is an open question, though recent advances in conformer generation with fixed compositions, such as adjoint Schrödinger bridge sampling~\cite{liu2025}, may provide useful insights.

Furthermore, several techniques, such as fake atoms and self-conditioning, were recently introduced in the unconditional FlowMol model~\cite{dunn2025-3} while this work was under development, and may further benefit the PropMolFlow conditional model. 
Despite these limitations and challenges, PropMolFlow demonstrates reliable generative modeling in chemical discovery and provides a strong foundation for future advancements.

%% file: Sections/methodology.tex
\newcommand{\p}{\mathbf{p}}
\newcommand{\bigsquarebracket}[1]{\left[#1\right]}
\newcommand{\bigcurlybracket}[1]{\left\{#1\right\}}
\newcommand{\bigbracket}[1]{\left(#1\right)}
\newcommand{\norm}[1]{\lVert #1 \rVert}
\newcommand{\aset}[1]{\{#1\}}
\newcommand{\nprod}{\prod_{i=1}^N}
\newcommand{\nsum}{\sum_{i=1}^N}

\subsection*{Joint flow matching\label{sec:joint_fm}}

Molecules are represented by fully-connected graphs $G$. 
Graph nodes encode the atomic types $A$, charges $C$ and positions $X$ of a molecule,  and graph edges include bond orders $E$. Ideally, new molecules would be generated by sampling directly from the probability distribution over valid molecular graphs $q(x)$, where $x$ is an arbitrary sample. However, this target distribution lacks a tractable analytical form and can only be approximated through a generative process. 
To this end, we adopt flow matching to learn an approximation of $q(x)$. 
This flow matching is parameterized by an SE(3) equivariant graph neural network built on top of FlowMol architecture~\cite{dunn2024-2}, details of which are provided in the subsequent \hyperref[sec:model_architecture]{`Model architecture'} section. 
In flow matching, the objective is to learn a time-dependent velocity field $u_t^\theta$ parameterized by a neural network $\theta$, which defines a probability path $p_t(x): [0, 1] \times x_0 \rightarrow x_1$ from a base distribution, $p_0(x)$ to a target distribution, $p_1(x)=q(x)$ (subscripts denote time dependence, for example, $p_t(x)$)
The flow induced by the velocity field is given by the ODE $\dot{x} = u_t \left(x\right)$, which describes the evolution of samples along the probability paths.
Designing a flow matching process thus involves two key steps: 1) specifying a probability path that satisfies the boundary conditions $x_0 \sim p_0$ and $x_1 \sim p_1$, and 2) training a neural network $\theta$ to approximate the velocity field that transports samples along this path.

In practice, while accurately modeling the exact velocity field is challenging, conditioning it on a random variable $z$ leads to a more tractable formulation. 
Importantly, optimizing the conditional velocity field is equivalent to optimizing the marginal velocity field~\cite{lipman2023}. The endpoint reformulation of \cite{dunn2024-1} reparametrizes the flow matching problem by focusing on learning a denoising function rather than the full time-dependent velocity field $u_t^\theta$. Specifically, the method learns a denoising function $x^\theta_{1|t}$ that predicts the final state $x_1$ from the intermediate state $x_t$. The training objective then minimizes the discrepancy between the predicted final sample and the true final sample:
\begin{equation}\label{eq:continuousfm}
\mathcal{L} = \mathbb{E}_{t, p_{t|1}(x_t|z), p_{z}} \bigsquarebracket{\norm{x^\theta_{1|t} - x_1}},
\end{equation}
where $z$ is a conditioning variable that can be either the final graph $x_1$ or a pair $z=(x_0, x_1)$ representing both the initial and final states.

Discrete graph features are handled via CTMC as developed by Campell \etal~\cite{campbell2024} and Gat \etal~\cite{gat2024}. 
In this setting, Eq.~\eqref{eq:continuousfm} is reformulated to minimize a cross-entropy loss over discrete variables:
\begin{equation}\label{eq:discretefm}
\mathcal{L}_{\rm{CE}} =  \mathbb{E}_{t, p_{t|1}(x_t|z), p_{z}} \bigsquarebracket{-\log p^\theta_{1|t} (x_1^i | x_t)}
\end{equation}

We factorize the conditional probability path of the molecular graph into the individual paths for each modality.  A joint flow matching process is then defined by learning a velocity field (or, equivalently, a denoiser)---parameterized by a GNN---that minimizes the total loss expressed as a weighted sum of the losses for each modality:
\begin{equation}\label{eq:total_loss}
\mathcal{L} = \eta_X \mathcal{L}_X +  \eta_A \mathcal{L}_A +  \eta_C \mathcal{L}_C + \eta_E \mathcal{L}_E  
\end{equation}
Previous studies suggest that atomic positions should have a higher weight, followed by bond orders, atomic charges, and atom types. 
The loss weights are chosen to be $(\eta_X, \eta_A, \eta_C, \eta_E) = (3.0, 0.4, 1.0, 2.0)$.
The factorization of the flow into its components allows for the flexible design of individual interpolants for each modality, as detailed below.

\paragraph{Interpolant for atomic positions.}
The atomic position flow is conditioned on the pair of initial and final states $z_X = (X_0, X_1)$, and is given by the linear interpolant
\begin{equation}\label{eq:x_interpolant}
X_t = \alpha_t X_0 + \beta_t X_1
\end{equation}
where $\alpha_0 = \beta_1 = 1$ and $\alpha_1=\beta_0=0$. 
The functions $\alpha_t$ and $\beta_t$, known as the interpolant schedule, control the rate of mixing between the base and target distributions.
A typical and effective choice of interpolant schedule for both molecule and materials generation is a linear schedule: $\alpha_t=(1-t)$ and $\beta_t=t$~\cite{dunn2024-2, hoellmer2025}.
Other variants of interpolants are discussed in detail in \cite{albergo2023, hoellmer2025}.
The base distribution for the atomic position flow is a standard Gaussian distribution $p_0(X) = \prod_{i=1}^N \mathcal{N} (X_0^i | \mathbf{0}, \mathbb{I}_3)$.
A denoising network is then trained to produce the final atomic positions $X^\theta_{1|t}$, minimizing the following loss:
\begin{equation}\label{eq:pos_loss}
\mathcal{L}_X=\mathbb{E}_{t, p_t(X_t|X_0, X_1), \pi(X_0, X_1)} \bigsquarebracket{\norm{X^\theta_{1|t} - X_1}}
\end{equation}
Where the joint distribution $\pi(X_0, X_1)$ defines the optimal transport coupling between $(X_0, X_1)$.
Details of the optimal transport formulation are provided in Supplementary Section \ref{sec:ot}.

\paragraph{Interpolant for atom type, charge, and bond order.}
Atom types, charge and bond orders are modeled using CTMC-based discrete flow matching~\cite{campbell2024}.
For each categorical variable, a `mask' token is added as an additional discrete state.
For example, consider atomic identities $A = \{A^1,A^2,\ldots,A^N\}$  where each $A^i$ takes values in the set $\aset{1, 2, \ldots, n_A, M}$. Here, $n_A$ is the number of predefined atom types and $M$ is the mask token.
The joint probability path over all atomic identities is factorized into independent contributions from each atom. The conditional probability path for atom $i$ is defined as:
\begin{equation}\label{eq:discrete_interpolant}
p_t(A^i_t|A_0, A_1) = \alpha_t \delta(A^i_t, A_1^i) + \beta_t \delta(A^i_t, M)
\end{equation}
where $\alpha_t$ and $\beta_t$ are the same interpolant schedules defined in Eq.~\eqref{eq:x_interpolant}. A linear interpolant is effective for categorical variables as well~\cite{dunn2024-2, hoellmer2025}.
Here, $\delta(i, j)$ denotes the Kronecker delta, which is 1 if $i=j$ and 0 otherwise. The base distribution corresponds to all atoms being initialized in the masked state.
This formulation implies that at time $t$, the atom identity $A^i_t$ has a probability $\alpha_t$ of being in its final state $A_1^i$ and a probability $\beta_t$ of being in the masked state $M$. 

In contrast to continuous variables, discrete flows are propagated via CTMC, where state transitions are stochastic rather than deterministic. The transition probability for atom $i$ over a short time interval $\Delta t$ can be expressed in terms of individual atomic contributions:
\begin{align}
p^i_{t + \Delta t}(j|A^i_t) &= \left\{\begin{array}{lcl}
 R^i(A^i_t, j)  &  \mbox{for} & j \neq A_t^i \\
 1 + R^i(A^i_t, A_t^i) & \mbox{for} & j = A_t^i 
\end{array}\right.  \\
 &= \delta(j, A_t^i) + R^i(A^i_t, j) \Delta t
 \label{eq:ctmc}
\end{align}
where $R(A^i_t, j)$ is the rate matrix specifying the transition probability from atom type $A^i_t$ to $j$~\cite{campbell2024}.  
To ensure proper normalization of the transition probabilities at time $t+\Delta t$, the rate matrix must satisfy $R^i(A^i_t, A^i_t) = - \sum_{j \neq A^i_t} R(A^i_t, j)$ so that the total probability mass remains normalized.

An additional stochastic term $\eta$ can be introduced into the rate matrix to form the modified rate matrix $R_t^\eta := R_t^* + \eta R_t^{\mathrm{DB}}$.
This new rate matrix generates the same marginal probability path if $R_t^{\mathrm{DB}}$ satisfies detailed balance (`DB') over the conditional probability path $p_{t|1}$.
By including the stochasticity term and choosing a linear interpolant ($\alpha_t=1-t$ and $\beta_t=t$), the rate matrix for atom $i$ becomes: 
\begin{equation}\label{eq:rate_matrix_with_eta}
R^i (A_t^i, j) = \frac{1+\eta t}{1-t} p^\theta_{1|t}(j|A_t^i) \delta (A_t^i, M) + \eta \bigbracket{1-\delta(A_t^i, M)}\delta(j, M)
\end{equation}
This formulation implies that if atom $i$ is in the masked state at time $t$, the probability of transitioning to atom type $j$ at time step $t + \Delta t$ is $\Delta t \frac{1 + \eta t}{1-t}$. Conversely, if atom $i$ is in an unmasked state, the probability of transitioning back to a masked state is $\eta \Delta t$. 
In Eq. \eqref{eq:rate_matrix_with_eta}, the rate matrix can diverge as $t \to 1$. To address this issue, we employ a time-dependent loss function that avoids division-by-zero~\cite{le2023}.

Additionally, low-temperature sampling---which rescales the prediction logits to sharpen the output distribution---is found to be critical for improving model performance~\cite{dunn2024-2}. Specifically, the denoiser network's output $p^\theta_{1|t}(A_1^i|A_t)$ is transformed as follows:
\begin{equation}\label{eq:output_conversion}
\widehat{p}^\theta_{1|t}(A_1^i|A_t) = \mathrm{softmax}\bigbracket{\nu^{-1} \log p^\theta_{1|t}(A_1^i|A_t)}
\end{equation}
where $\nu$ is a hyperparameter. The modified probabilities $\widehat{p}^\theta_{1|t}(A_1^i|A_t)$ are then employed in the discrete loss function defined in Eq.~\eqref{eq:discretefm}.
When training PropMolFlow, we set $\eta=10$ and $\nu=0.05$.
During inference, we observe that turning off the stochastic term leads to better performance.
To ensure that the generated molecules satisfy the required symmetry constraints, the joint flow matching process is constructed with an invariant base distribution and an equivariant transition probability path parameterized by the SE(3)-GVP architecture. A self-contained formal derivation of probability invariance and equivariant generative process is provided in respective Supplementary Section~\ref{sec:equivariantfm} and Supplementary Section~\ref{sec:se3_gvp}.

\subsection*{Model architecture}\label{sec:model_architecture}
We utilize the FlowMol architecture, which is implemented with PyTorch and Deep Graph Library~\cite{dunn2024-2}.
Molecule updates are achieved through layers comprising Geometric Vector Perceptrons (GVP).
Within each GVP, the molecule graph passes through a sequential steps of node feature update, node position update and edge feature update. 
Each node $i$ consists of a position $x_i \in \mathbb{R}^3$, scalar features $s_i \in \mathbb{R}^d$, and vector features $v_i \in \mathbb{R}^{c\times 3}$.
The scalar feature is a concatenation of atom type and charge vectors; that is, $s_i:= [a_i: c_i]$ where `:' defines a concatenation operation.
Vector features are initialized at zeros, and despite not existing in the final outputs, their updates employ the vector cross-products that is not equivariant to reflections, hence SE(3) equivariant~\cite{dunn2024-1}. 
Each edge feature corresponds to the bond order and the permutation invariance of the scaler bond order is ensured by taking the sum of learned bond features from $i \to j$ and $j \to i$;\ie, $\widehat{e}^{ij} = \mathrm{MLP} (e_{ij} + e_{ji})$.

\paragraph{Node feature update.} The node feature uses graph convolution to update node scalar and vector features $s_i, v_i$. It follows two steps: the first step generates scalar $m_{i \to j}^{(s)}$ and vector messages $m_{i \to j}^{(v)}$ by a function $\psi_M$ which is a chain of two GVPs:

\begin{equation}\label{eq:nfu_message_generation}
m_{i \to j}^{(s)}, m_{i \to j}^{(v)} = \psi_M \bigbracket{\bigsquarebracket{s_i^{(l)}: e_{ij}^{(l)} : d_{ij}^{(l)}}, 
\bigsquarebracket{v_i: \frac{x_i^{(l)}- x_j^{(l)}}{d_{ij}^{(l)}}} }
\end{equation}
Where $d^{(l)}_{ij}$ is the distance between nodes $i$ and $j$ at update block $l$.
In the FlowMol implementation, the distance $d_{ij}$ is replaced with a radial basis distance embedding before being fed into the GVPs or MLPs.
Following the message generation step, an update of node scalar and vector features can be accomplished by aggregating generated messages:

\begin{equation}\label{eq:nfu_node_update}
s_i^{(l+1)}, v_i^{(l+1)} = \mathrm{LN} \bigbracket{[s_i^{(l)}, v_i^{(l)}] + \psi_N \bigbracket{\frac{1}{|\mathcal{N}(i)|} \sum_{j \in \mathcal{N}(i)}} \bigsquarebracket{m_{j \to i}^{(s)}, m_{j \to i}^{(v)}} } 
\end{equation}
Where $LN$ stands for LayerNormalization operation, $\psi_N$ is a chain of three GVPs.

\paragraph{Node position update.} NPU block updates node positions by taking node-wise operations on the updated node scalar and vector features:

\begin{equation}\label{npu}
x_i^{(l+1)} = x_i^{(l)} + \psi_P \bigbracket{s_i^{(l+1)}, v_i^{(l+1)}}
\end{equation}
Where $\psi_P$ is a chain of three GVPs in which the final output has 1 vector and 0 scalar features. 

\paragraph{Edge feature update.} Edge features are updated by edge-wise operations that take the updated node scalar features and node distance as the inputs:
\begin{equation}\label{eq:edge_update}
e_{ij}^{(l+1)} = \mathrm{LN} \bigbracket{e_{ij}^{(l)} + \mathrm{MLP} \bigbracket{s_i^{(l+1)}, s_j^{(l+1)}, d_{ij}^{(l+1)}}} 
\end{equation}
Models are trained using the settings detailed in the \hyperref[sec:training_details]{`Hyperparameters and training settings'} section.

\subsection*{Property embedding operations\label{sec:prop_emb}}

Conditional generation can be framed as sampling a property-conditional target distribution $p_1(x|k)$ over molecular graphs, where $k$ is a property.
Previous approaches to conditional generation concatenate a property value to the node features of a molecular graph, training the model jointly on the property and the graph~\cite{hoogeboom2022, xu2023, song2023}.
As a result, the learned denoising process is conditioned on the property.
This approach requires modifying the input dimension of the denoiser to include the additional property value, which is subsequently discarded in the final output.
However, directly appending the property to the node features may oversimplify the interaction between property values and molecular graphs.
As shown in Figure \ref{fig:propmolflow_overview}, our implementation maps a scalar property, $k$,  to a high-dimensional property embedding space $P=\phi_{\mathrm{prop}}(k)$ using a MLP, $\phi_{\mathrm{prop}}$.
The dimensionality of the property embedding matches that of the node scalar features, allowing seamless integration without requiring additional dimensional matching. To encode interactions between the property embedding and node scalar features in molecualr graphs, we propose five distinct operations, including `Concatenation', `Sum', `Multiply', `Concatenate + Sum', and `Concatenate + Multiply'.
Depending on the operation, if the combined feature vector $[A, C]$ maintains its original dimensionality after interacting with the property embedding, no further processing is applied; otherwise, an MLP is used to project it back to the original dimensionality.
Since the property embedding interacts exclusively with node scalar features, leaving position vectors unchanged, the resulting property-conditioned graph retains the SE(3) equivariance. Moreover, the loss function preserves its original form, and the trained SE(3) GNN defines a valid conditional generative process.

Specifically, in the `Sum' operation, the property embedding is simply added to the node scalar features.
The `Multiply' operation performs an element-wise Hadamard product `$\odot$'.
To ensure the multiplicative factors remain within a bounded range, the property embedding is passed through a sigmoid function and then shifted by 0.5, resulting in multipliers in the interval $[0.5, 1.5]$. Both `Sum' and `Multiply' preserve the original feature dimensionality.
In contrast, operations involving `Concatenate' increase the dimensionality of the node scalar features, and an MLP transformation $\varphi_\theta$ is required to project the combined features back to the original dimension.
For the composite operations, `Concatenate + Sum' and `Concatenate + Multiply', the property embedding is first concatenated to the node features, followed by the respective arithmetic operation. In both cases, the final output has the desired dimensionality, either by design or via the subsequent MLP transformation.

\subsection*{Gaussian expansion\label{sec:ge}}
We incorporate a fixed, non-trainable Gaussian expansion layer to enrich the property representations prior to being mapped to a property embedding via an MLP $\phi_{\mathrm{GE}}$~\cite{gebauer2022}.
The Gaussian expansion maps a scalar property value into a fixed-length vector using a set of Gaussian basis functions. Given a property $k$ with value $\tau_k$, the expanded representation takes the form:
\begin{equation}\label{eq:gaussian_expansion}
f_{\mathrm{k}} = \phi_{\mathrm{GE}} \bigbracket{\bigsquarebracket{\exp \bigbracket{-\frac{(\tau_\mathrm{k}-(\tau_{\mathrm{min}} + n_g d))^2}{2 d^2}}}_{0 \leq n_g \leq \frac{\tau_{\mathrm{max}} - \tau_{\mathrm{min}}}{d}}}
\end{equation}
where $\tau_\mathrm{min}$ and $\tau_\mathrm{max}$ are the minimum and maximum property values, $d$ is the uniform  spacing between the Gaussian centers, and $n_g$ indexes the basis functions.  The output vector encodes the proximity of the input $\tau_k$ to each Gaussian center, producing a smooth, localized representation that is refined by the MLP $\phi_{\mathrm{GE}}$.

Figure \ref{fig:propmolflow_overview}d shows an example of a Gaussian expansion for the HOMO-LUMO gap.
In this example, five Gaussian basis functions ($n_g =5$) are used, with centers evenly spaced between the minimum and maximum gap values in the QM9 dataset: $\tau_\mathrm{min}=0.0246$ and $\tau_\mathrm{max}=0.6221$ Hartree. 
The width of each Gaussian is set by the spacing parameter $d = (\tau_\mathrm{max} - \tau_\mathrm{min}) /  (n_g - 1)$.
The top panel of Figure \ref{fig:propmolflow_overview}d shows the five Gaussian basis functions, with their centers marked by vertical dashed gray lines.
In the bottom panel, we pick two example property values, namely 0.1, and 0.5 Hartree, and show the expanded feature vectors as bar plots indicating the activation of each Gaussian basis function.
For instance, if the gap is 0.1 Hartree,  the first two Gaussian functions (centered closest to 0.1) have higher responses than the remaining functions, whose centers are farther away. Similar activation patterns are observed for the other two property values.

\subsection*{Inference}\label{sec:inference}

\paragraph{ID tasks.} To generate paired samples of property values and numbers of atoms, we employed a two-stage sampling procedure. First, atom counts were sampled according to their empirical distribution in the training dataset. Specifically, the frequency of each unique atom count was computed and used to construct a categorical probability distribution, from which atom counts were drawn.

Conditional on a given atom count, the associated property values were sampled using a histogram-based approximation of their empirical distribution. For each unique atom count, the observed property values were discretized into a fixed number of bins like 1000. The counts within each bin were normalized to form a categorical distribution. During sampling, a bin index was first drawn from this categorical distribution, after which a continuous property value was assigned by uniformly sampling within the range of the selected bin.

This approach yields paired samples of atom counts and property values that respect both the marginal distribution of atom counts and the conditional distribution of properties given the atom count. 
To make a fair comparison across different property embedding methods, the same sampled target values and atom counts were used.

\paragraph{OOD tasks.} The property values in the OOD tasks are chosen as the 99$^{\rm{th}}$ quantile of the QM9 training data and the numbers of atoms are selected for the property ranges bounded by the 97th quantile and the maximum (100th quatile) property value. 
During inference, PropMolFlow models use Euler's method with 100 evenly spaced time steps to integrate the learned velocity field. 

\subsection*{Datasets, baselines and evaluation metrics}\label{sec:data_and_metrics}

\subsubsection*{Datasets}\label{sec:dataset}

We trained PropMolFlow on the QM9 data with explicit hydrogen atoms~\cite{ramakrishnan2014quantum, wu2018moleculenet}.
We did not evaluate PropMolFlow on the GEOM-Drugs dataset, which is commonly used for unconditional generation and includes larger molecules, because it only provides quantum-mechanical energy values and lacks other relevant properties~\cite{axelrod2022}.
The original QM9 data contains 133,885 molecules, and each molecule contains up to 9 heavy atoms and consists of 3--29 atoms (average: 18 atoms), composed of up to 5 elements: C, N, O, F, and H.
Molecules are fully optimized and molecular properties are obtained with DFT at the level of B3LYP/6-31G(2df,p)~\cite{becke1993, lyp1988}.
For property-guided generation, we considered six molecule properties: polarizability ($\alpha$), HOMO-LUMO gap, HOMO energy, LUMO energy, dipole moment ($\mu$), and heat capacity ($C_v$). Precise definitions of each property can be found in Supplementary Section \ref{sec:more_details}.
Upon inspection, we discovered numerous inconsistencies: invalid bond orders and non-zero net charges in roughly 30,000 molecules, despite QM9's requirement for charge-neutral, closed-shell valency~\cite{ramakrishnan2014quantum}.
We corrected these discrepancies by reassigning bond orders or charges to enforce valency-charge consistency for the vast majority of entries and used the corrected data to train our PropMolFlow models.
After the correction, we improved the structural validity of generated molecules substantially (Supplementary Table~\ref{tab:data_and_metric_compare}). 
Previous works that used bond orders overlooked this issue~\cite{vignac2023, dunn2024-2}; instead, they modified evaluation metrics for models trained on the problematic SDF data against the one used by Hoogeboom et al.~\cite{hoogeboom2022} to tolerate chemically inconsistent bond--charge combinations, hence inflating molecule stability estimates contrary to QM9's intended constraints.
A detailed comparison between models using previous problematic data and the corrected data, together with the comparison between previous stability metrics and the revised ones, can be found in Supplementary Table \ref{tab:data_and_metric_compare}. 
A detailed description of our procedure and the alignment with the original QM9 XYZ data is provided in Supplemental Section \ref{sec:qm9_data_fix}.
The revised QM9 SDF data is provided on Zenodo~\cite{zenodo_propmolflow}.
We used RDKit's sanitization function~\cite{rdkit} to filter chemically valid molecules on the QM9 SDF file with bond and charge fix, resulting in a curated set of 133k molecules. After random shuffling, we split the dataset into 100k training, 20k validation, and 13k test samples.
PropMolFlow models and GVP property predictors are trained on disjoint 50k and 50k datasets from the 100k training data. 

\subsubsection*{Baselines}\label{sec:baselines}

We compare PropMolFlow against several recent property-guided molecular generation methods based on diffusion models and flow matching, including EEGSDE, GCDM, GeoLDM, JODO, and EquiFM~\cite{bao2023,  morehead2024, xu2023, huang2024, song2023}.
EEGSDE (Equivariant Energy Guided Stochastic Differential Equations) use an energy function to guide the diffusion for property-guided generation~\cite{bao2023}.
EquiFM is the first equivariant flow matching model for 3D molecule generation~\cite{song2023}.
GeoLDM represents the first latent diffusion model for molecular geometry generation~\cite{xu2023}.
GCDM constructs an SE(3) equivariant geometry-complete diffusion model using geometry-complete perceptrons for molecule generation~\cite{morehead2024, moret2023}.
JODO uses a diffusion graph transformer for joint generation of 2D molecular graphs and 3D geometries.
EEGSDE, EquiFM, GeoLDM, and GCDM do not include bond orders in their molecular graphs, while JODO incorporates both bond existence and bond orders. 
All five baseline models omit atomic charges in their conditional generation, in contrast to PropMolFlow, which can generate molecules with non-zero atomic charges. 

In this work, we focus exclusively on conditional generation results; unconditional generation performance is reported in earlier studies.
Language-model-based approaches are excluded from this comparison, as state-of-the-art models in that category rely on fundamentally different molecule representations and training data regimes.
For EEGSDE, we benchmark PropMolFlow against the canonical conditional model with a scaling factor of 1 ($s=1$)~\cite{bao2023}. 
Although EEGSDE provides DFT-based validation for five out of six properties, it does so on only 100 generated molecules,  which limits the statistical significance of those results.
Because EEGSDE, JODO and GCDM released their model checkpoints, we sampled molecules from those and computed all structural validity metrics on the resulting structures; for GeoLDM, we used the checkpoints bundled with GCDM.
EquiFM checkpoints were unavailable, so we trained its conditional models by ourselves following information in~\cite{song2023} (training details in Supplementary Section \ref{sec:more_details}).

Three intuitive baselines are included for comparison in the ID task. 
The ``Random (Upper-bound)'' intuitive baseline fully shuffles property values, thereby removing any correlation between molecular structures and properties, and serves as an upper bound on error.
The ``\# Atoms'' baseline uses only the number of atoms as a predictor for molecular properties, capturing coarse size-dependent trends.
It is implemented as a simple neural network with one hidden layer~\cite{hoogeboom2022}.
The ``QM9 (Lower-bound)'' baseline corresponds to a pretrained EGNN regressor provided by Hoogeboom \etal~\cite{hoogeboom2022}, trained directly on QM9 and serving as a lower bound on achievable error.
Improvement over the ``Random'' baseline indicates that the conditional generation effectively integrates property information.
Surpassing the ``\# Atoms'' baseline suggests that the generative model captures structural features beyond simple atom count when generating new molecules.

\subsubsection*{Evaluation metrics}\label{sec:evaluation_metrics}

\paragraph{Structural validity metrics.} Our comparison includes both structural-validity, inference-efficiency and property-specific evaluation metrics.
Structural validity metrics include atomic stability, molecule stability, RDKit validity, `uniqueness and validity', PoseBusters validity, and closed-shell ratio.
Atom stability is given by the proportion of atoms with correct valency. 
To account for atomic charges, an atom is considered to have the correct valency if its formal charge balances its explicit valency, as defined in RDKit~\cite{rdkit}.
For instance, a nitrogen atom should carry a +1 formal charge if it has a valency of 4.
Molecule stability defines the proportion of molecules in which all atoms are stable and the molecule is charge neutral (zero net formal charges).
RDKit validity refers to the ratios of molecules that pass RDKit's sanitization check.
`Uniqueness and validity' is defined as the proportion of molecules that are RDKit valid and unique in their SMILES representation.
PoseBusters validity corresponds to the fraction of molecules that pass many \textit{de novo} chemical and structural validation tests, such as all atom connectivity, valid bond lengths and angles, and absence of internal clashes~\cite{buttenschoen2024}.
We used the first 10 columns of the PoseBusters outputs in CSV formats as the criteria for results shown in Fig.~\ref{fig:standard_metrics} and Supplementary Table~\ref{tab:standard_metrics}.
A molecule is closed-shell valid if it has an even number of valence electrons. Since all molecules in QM9 are closed-shell, any generated open-shell species are considered invalid.
Except for the closed-shell ratio, all structural validity metrics depend on bond-order information, which baseline models (other than JODO) do not output.
We therefore assigned bond orders using the distance-based, optimized cutoffs introduced by Hoogeboom~\cite{hoogeboom2022}. Although recent work~\cite{vignac2023} suggests that the bond orders can probably be optimized through the OpenBabel program~\cite{O'Boyle2011}, we did not apply such refinements to the baseline samples.
To assess the chemical similarity between generated molecules and a reference training set, we employed a fingerprint-based approach using RDKit. 
For each molecule, a Morgan fingerprint (circular fingerprint) was computed with a radius of 2 and a 2048-bit representation.
Pairwise similarity between each generated molecule and all training molecules was computed using the Tanimoto similarity metric on their respective Morgan fingerprints. For each generated molecule, the maximum similarity score and the corresponding most similar training molecule were recorded.
Lastly, inference efficiency is given by the wall-clock time required to generate 10,000 molecules.

{Property metrics.} PropMolFlow is directly compared to prior methods on the ID tasks.
In this task, a conditional generative model was first used to generate molecules conditioned jointly on property values and atom counts sampled from the same distribution as the training data (for example, QM9). These sampled property values and atom counts served as the input conditions. 
A separate property predictor/regressor was then applied to estimate the properties of the generated molecules.
In PropMolFlow,  we used an SE(3) GVP network for molecule generation and, correspondingly, trained a separate GVP regressor for each of the six molecular properties.
10000 molecules are generated to evaluated the performance of conditional generation. OOD results are not directly comparable to baseline models, but we validate our results with DFT calculations to assess physical fidelity. 

\paragraph{RMSD and property sensitivity.} Given the same chemical compositions and element-type ordering, the RMSD between any two molecules (M$_0$ and M$_1$) is the minimum distance between these two structures considering the translational and rotational symmetry operations, which can be expressed as:
\begin{equation}\label{eq:rmsd}
\mathrm{RMSD}(\rm{M}_0, \rm{M}_1) = \min_{P \in SE(3)} \{d\left(M_0, P(M_1)\right)\}
\end{equation}
Where SE(3) represents the group of operations that respect the translational and rotational symmetry.
To estimate the property sensitivity to structural relaxation, we selected the structures whose RMSDs are larger than 0.03 \AA---RMSDs lower than this threshold are considered very close to the relaxed structures, and in Eq. \eqref{eq:normalized_sensitivity}, we define the normalized property sensitivity `$\chi_{\rm{DFT}}$' to allow comparisons between different properties.
\begin{equation}\label{eq:normalized_sensitivity}
\chi_{\rm{DFT}} = \frac{\vert{\delta q_{\rm{DFT}}\vert}}{(q_{0.99} - q_{0.01}) \cdot \rm{RMSD}}
\end{equation}
where $\delta q_{\rm{DFT}}$ is the change of DFT-calculated property upon relaxation, and the normalizer ($q_{0.99} - q_{0.01}$) is the difference between the 99th quantile and the 1th quantile of the property distribution of QM9 training data.
This range is chosen over the min-max range because it is robust to outliers.

\subsection*{Computational settings}\label{sec:computational_setting}

\subsubsection*{GVP regressor details}\label{sec:gvp_details} 
To be self-consistent, we trained property regressors using Graph Neural Networks based on GVPs. 
GVP property predictors are trained on a disjoint dataset versus that used for training PropMolFlow models, hence avoiding data leakage between the generative models and property prediction models.
We appended an MLP layer that takes the final node scalar features as input to predict the target property.
The parameters of GVP regressors are optimized by minimizing a mean squared error loss function.
Separate models were trained for each of the six molecular properties.

\subsubsection*{DFT}\label{sec:dft_details}
All DFT calculations were performed using the Gaussian 16 package (Gaussian 16, Revision C.01)~\cite{g16}. The B3LYP hybrid functional~\cite{becke1993, lyp1988} together with the 6-31G(2df,p) basis set was employed for single-point calculations, and geometric optimizations.
Single-point calculations were used unless otherwise specified. Structural relaxations were performed for the property $C_v$ in most cases, except in the interpolation study. Relaxations were performed to evaluate changes in both properties and geometries, as shown in Fig.~\ref{fig:target_gvp_dft}. DFT calculations that failed to converge were discarded.
We released the DFT-evaluated molecules in both the ID and OOD tasks with and without structural optimization at the Zenodo repository~\cite{zenodo_propmolflow}.

\subsubsection*{Molecule filtering}\label{sec:molecule_filtering}
We selected 1000 molecules out of 10000 samples structures in the ID task or use the directly generated 1000 structures in the OOD task for DFT evaluations.
multi-step filtering is essential to ensure that the structures are chemically valid and suitable for downstream applications.
We filtered the 1000 molecules by five criteria: Molecule stability, RDKit validity, PoseBusters validity, closed-shell validity, and multi-fragment check.
The multi-fragment check filters out molecules that with more than one disconnected fragment.
For molecules passing all filters, property values are evaluated using both a GVP property predictor and DFT calculations. 
We also filtered out molecules whose DFT calculations did not converge, resulting in no less than 90\% of and 78--94\% of the molecules remaining for the subsequent analyses in the respective ID and OOD tasks (Supplementary Table \ref{tab:no_sp_dft}).

\subsubsection*{Hyperparameters and training details}\label{sec:training_details}
PropMolFlow models on the QM9 dataset were trained with 8 molecule update blocks. 
Atoms contain 256 hidden scalar features and 16 hidden vector features. 
Edges contain 128 hidden features.
QM9 PropMolFlow models were trained with 2000 epochs.
For each property embedding method, 6 model checkpoints with lowest validation losses were saved and the top-3 model checkpoints were deposited to the Zenodo repository~\cite{zenodo_propmolflow}. 
Inference for PropMolFlow models and all baseline models used a single NVIDIA A100-SXM4 graphic card with 80GB of memory, with a batch size of 128.
Training of PropMolFlow models use both Nvidia A100-SXM4 and 2080ti graphic cards.
PropMolFlow models can be trained in 2--3 days with A100 and 4--5 days with 2080Ti cards, and EquiFM models were trained with 2500 epochs on A100, which took around 3.5 days.